\documentclass[conference]{IEEEtran}
\IEEEoverridecommandlockouts
\usepackage{cite}
\usepackage{amsmath,amssymb,amsfonts}
\usepackage{algorithmic}
\usepackage{graphicx}
\usepackage{textcomp}
\usepackage{xcolor}
\usepackage{subfigure}
\usepackage{booktabs}

\def\BibTeX{{\rm B\kern-.05em{\sc i\kern-.025em b}\kern-.08em
    T\kern-.1667em\lower.7ex\hbox{E}\kern-.125emX}}
\newcommand\blfootnote[1]{%
  \begingroup
  \renewcommand\thefootnote{}\footnote{#1}%
  \addtocounter{footnote}{-1}%
  \endgroup
}

\begin{document}

\title{A Cross-Framework Study of Temporal Information Buffering Strategies for Learned Video Compression\\
}

\author{%
\IEEEauthorblockN{
Kuan-Wei Ho\IEEEauthorrefmark{1} 
\quad Yi-Hsin Chen\IEEEauthorrefmark{1} 
\quad Martin Benjak\IEEEauthorrefmark{2}
\quad Jörn Ostermann\IEEEauthorrefmark{2}
\quad Wen-Hsiao Peng\IEEEauthorrefmark{1}
}
\IEEEauthorblockA{%
\IEEEauthorrefmark{1}National Yang Ming Chiao Tung University, Taiwan \quad
\IEEEauthorrefmark{2}Leibniz Universität Hannover, Germany
}
}

\maketitle

\begin{abstract}
Recent advances in learned video codecs have demonstrated remarkable compression efficiency. Two fundamental design aspects are critical: the choice of inter-frame coding framework and the temporal information propagation strategy. Inter-frame coding frameworks include residual coding, conditional coding, conditional residual coding, and masked conditional residual coding, each with distinct mechanisms for utilizing temporal predictions. Temporal propagation methods can be categorized as explicit, implicit, or hybrid buffering, differing in how past decoded information is stored and used. However, a comprehensive study covering all possible combinations is still lacking. This work systematically evaluates the impact of explicit, implicit, and hybrid buffering on coding performance across four inter-frame coding frameworks under a unified experimental setup, providing a thorough understanding of their effectiveness.

\end{abstract}

\begin{IEEEkeywords}
Learned video compression, hybrid temporal information buffering, residual coding, conditional coding, conditional residual coding, masked conditional residual coding.
\end{IEEEkeywords}

\vspace{-0.4cm}
\blfootnote{This work is supported by the National Science and Technology Council (NSTC), Taiwan, under Grants NSTC 113-2634-F-A49-007- and NSTC 114-2221-E-A49-035-MY3. We thank the National Center for High-performance Computing (NCHC) for providing computational and storage resources.}

\section{Introduction}
\label{sec:intro}

\begin{figure}[t]
    \centering
    \includegraphics[width=0.81\linewidth, trim=0 10 30 15, clip]{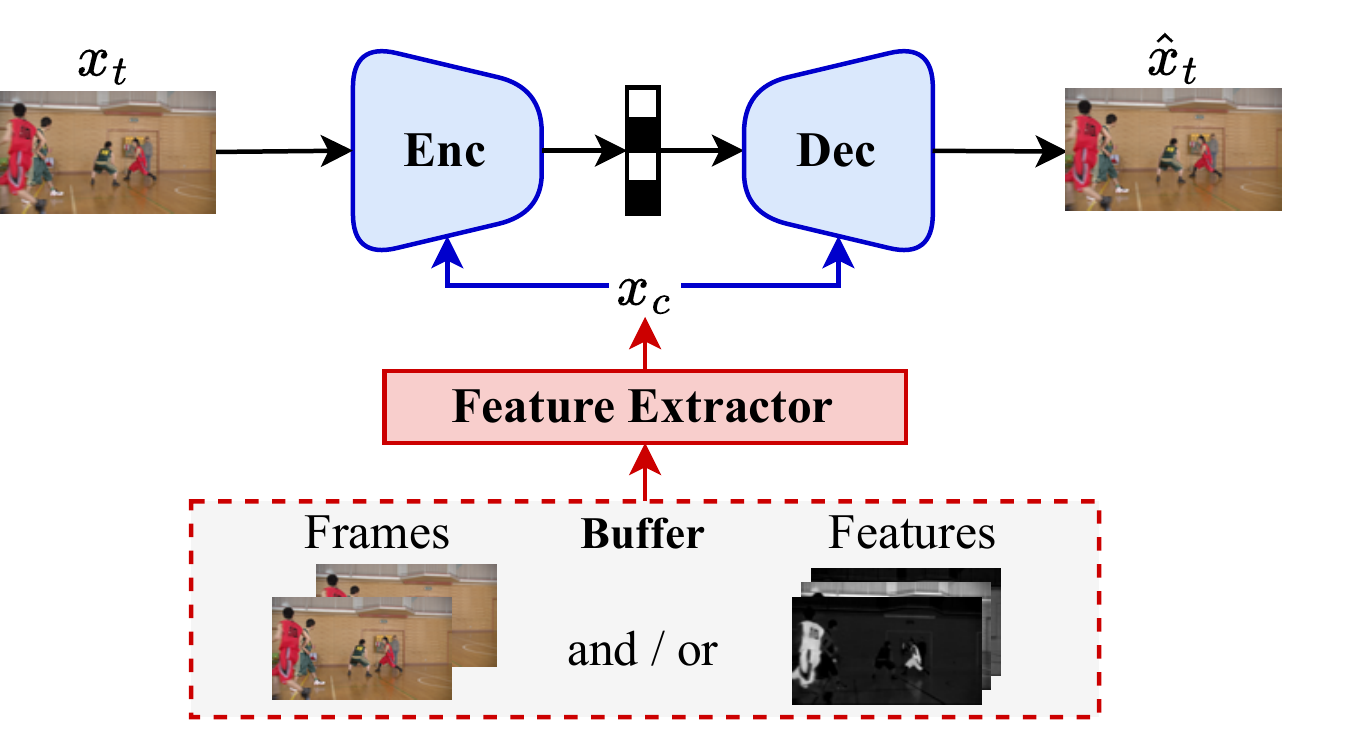}
    \vspace{-3mm}
    \caption{Two fundamental aspects in the design of learned video codecs: the inter-frame coding framework (highlighted in blue) and the mechanism of temporal information propagation (highlighted in red).}
    \label{fig:Teaser}
    \vspace{-0.4cm}
\end{figure}

Learned video codecs~\cite{dcvc_dc, dcvc_fm} achieve promising coding performance by leveraging neural networks to model temporal dependencies across frames and perform compression using neural inter-frame codecs. As illustrated in Fig.~\ref{fig:Teaser}, two fundamental aspects are among the most critical in their development: the design of the inter-frame coding framework (highlighted in blue) and the mechanism of temporal information propagation (highlighted in red).

The design of the inter-frame coding framework focuses on how the temporal prediction $x_c$ can be effectively utilized to assist the coding of the current frame $x_t$. Recent inter-frame coding frameworks can be categorized into four types: residual coding, conditional coding, conditional residual coding, and masked conditional residual coding. Residual coding~\cite{dvc, ssf} encodes the residue $x_t - x_c$, while conditional coding~\cite{dcvc_dc, dcvc_fm} enables nonlinear use of $x_c$ by directly encoding $x_t$, with $x_c$ serving as a condition signal for both the inter-frame encoder and decoder. While conditional coding generally outperforms residual coding, it can encounter an information bottleneck issue~\cite{pcs22}, where information carried by $x_c$ is potentially lost during neural network processing of $x_c$. To alleviate this issue, conditional residual coding~\cite{crc} combines residual coding and conditional coding by encoding the residue $x_t-x_c$ with a conditional inter-frame codec. Under the assumption that $H(x_t-x_c) \leq H(x_t)$, conditional residual coding achieves better coding performance. However, this assumption may not hold in local regions with dis-occlusion or unreliable motion estimates. To mitigate the sub-optimal performance caused by such violations, masked conditional residual coding~\cite{maskcrt} employs a soft mask to blend conditional coding and conditional residual coding at the pixel level.

\begin{figure*}[t]
    \centering
    \includegraphics[width=0.94\linewidth, trim=38 0 39 9, clip]{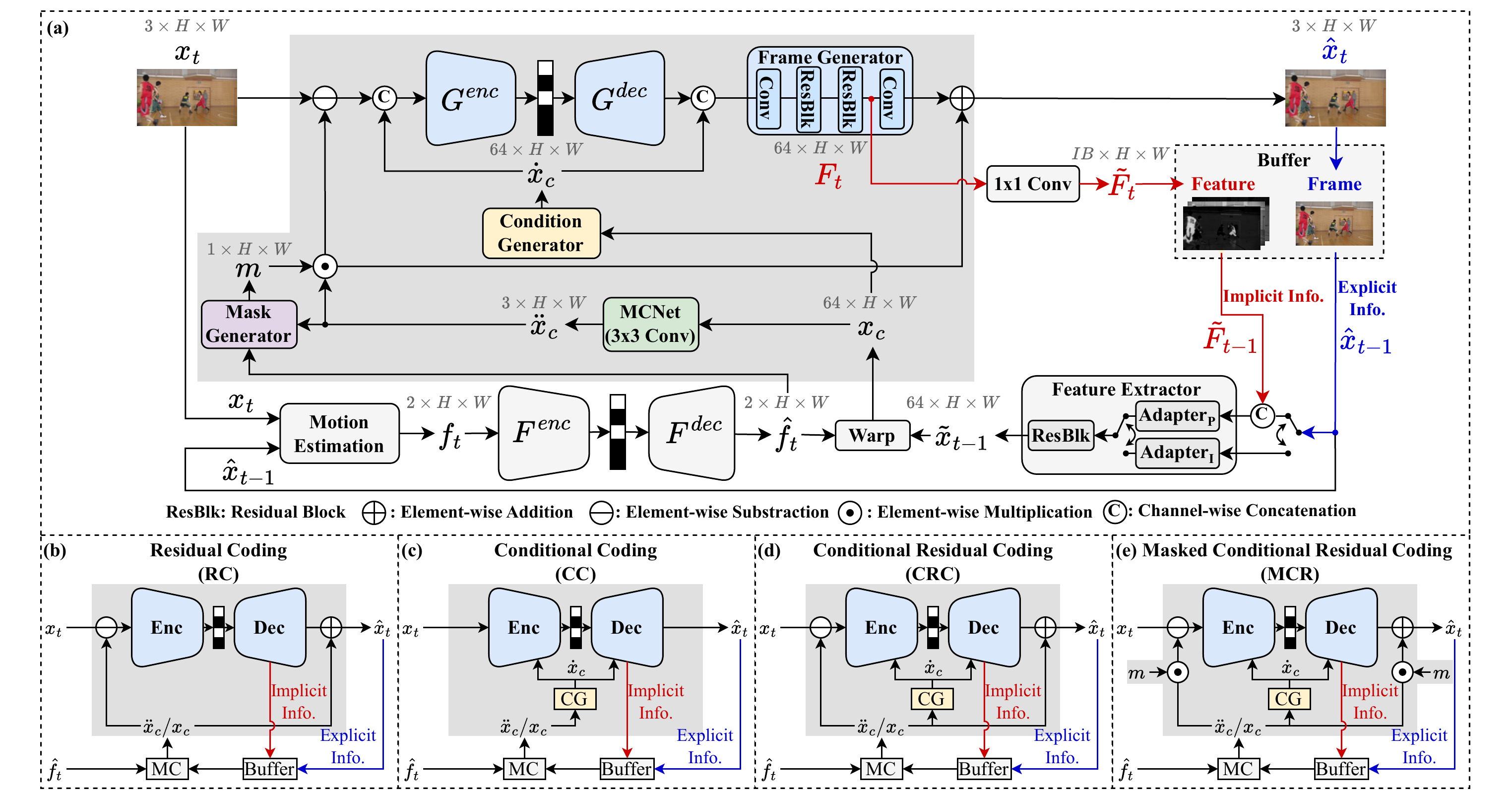}
    \vspace{-3mm}
    \caption{System Overview. (a): the architecture of the masked conditional residual coding framework with hybrid buffering used in our study. (b)–(e):  the architectural variations across different coding frameworks.}
    \vspace{-5mm}
    \label{fig:Overview}
\end{figure*}

\begin{table}[t]
\caption{Evaluated combinations of coding frameworks and buffering strategies in prior works with cross-variant discussion. This work evaluates all combinations.}
\vspace{-2mm}
\label{table:related}
\centering
\setlength{\tabcolsep}{6pt}
\begin{tabular}{l|c|c|c|c}
\toprule
\begin{tabular}[c]{@{}c@{}} Buffering\\Strategies  \end{tabular}  & RC & CC & CRC & MCR  \\
\midrule
Explicit &  ~\cite{crc,maskcrt}  &  ~\cite{crc,maskcrt,mmsp24,icme25}  & ~\cite{crc,maskcrt,mmsp24,icme25} & ~\cite{maskcrt,mmsp24,hytip}  \\
Implicit & -  & -  & ~\cite{icme25} & ~\cite{hytip}   \\
Hybrid  & -  &  ~\cite{icme25}  & ~\cite{icme25} & ~\cite{hytip}   \\
\bottomrule
\end{tabular}
\vspace{-7mm}
\end{table}

Temporal propagation in learned video codecs focuses on effectively utilizing information produced during previous decoding process to formulate the temporal prediction $x_c$. Based on the nature of the buffered signals, recent works can be categorized into three buffering strategies: explicit, implicit, and hybrid. Explicit buffering~\cite{dvc, ssf, dcvc, canfvc, pcs22, crc, maskcrt, mmsp24} stores the previously decoded frame $\hat{x}_{t-1}$ as a reference, relying on its strong correlation with the current coding frame $x_t$. However, the decoded frames must simultaneously approximate the input accurately and summarize useful information from the past. Implicit buffering~\cite{dcvc_dc, dcvc_fm} instead stores a large number of high-dimensional intermediate features during the decoding process, avoiding this dual constraint but often resulting in non-compact representations. Some models~\cite{dcvc_dc, dcvc_fm} store more than 48 high-resolution feature maps, leading to high memory bandwidth demands during decoding. Hybrid buffering~\cite{icme25, hytip} combines explicit and implicit mechanisms by storing $\hat{x}_{t-1}$ as the primary reference along with a few compact complementary features, reducing overall buffer size requirements.

As shown in Fig.~\ref{fig:Teaser}, the inter-frame coding framework and temporal propagation mechanism are two orthogonal design choices and can be combined flexibly. Table~\ref{table:related} shows that while some studies~\cite{crc,maskcrt,mmsp24,icme25,hytip} explore partial combinations of different coding frameworks and temporal propagation strategies, none provides a comprehensive analysis of all four inter-frame coding frameworks across the three temporal buffering methods. In this work, we systematically investigate how explicit, implicit, and hybrid temporal buffering strategies impact coding performance across various inter-frame coding frameworks using a unified backbone to ensure a fair comparison.

\section{Temporal Buffering Strategies \\across Coding Frameworks}
\label{sec:method}

Fig.~\ref{fig:Overview}(a) illustrates the architecture of the masked conditional residual coding (MCR) framework with hybrid buffering used in our study, which is slightly adapted from \cite{dcvc, mmsp24, icme25}. It consists of (1) a motion estimation network to estimate the motion between the reference frame $\hat{x}_{t-1}$ and the current coding frame $x_t$; (2) a motion codec $\{ F^{enc}, F^{dec} \}$ to encode the estimated flow map $f_t$; (3) a feature extractor, comprising a switchable adapter (either a $3\times3$ convolution for Adapter$_I$ or a $1\times1$ convolution for Adapter$_P$) followed by a residual block, to extract features from explicit and implicit references; (4) an inter-frame coding module, highlighted with a gray background (including MCNet, Mask Generator, Condition Generator, an inter-frame codec $\{ G^{enc}, G^{dec} \}$, and Frame Generator) to encode the coding frame $x_t$ using the warped feature-domain temporal prediction $x_c$ as reference; and (5) a $1 \times 1$ convolution to adapt the channel size of the intermediate features during decoding, $F_t$, for implicit buffering. In our study, we adjust the channel size $IB$ of this convolutional layer to examine the impact of buffer size on coding performance.

To ensure a fair comparison among different buffering strategies and inter-frame coding framework variants, we minimally modify this framework according to each variant.

\begin{table*}[t!]
\caption{BD-rate (\%) comparison with HM~\cite{hevc} in terms of PSNR-RGB, using residual coding with explicit buffering as the anchor.}
\vspace{-2mm}
\label{table:Main_RD_BT709}
\centering
\setlength{\tabcolsep}{9pt}
\begin{tabular}{l|ccccccc|c}
\toprule
Variants (Buffer size)  & UVG     & HEVC-B  & HEVC-C  & HEVC-D  & HEVC-E  & HEVC-RGB & MCL-JCV & Average \\
\midrule 
RC, Explicit (3+0)      &  0      &  0      &  0      &  0      &  0      &  0      &  0      &  0     \\
RC, Implicit (0+67)     &   5.6   &  15.3   &  20.2   &  29.6   &  55.2   &  18.1   &  11.8   &  22.3  \\
RC, Implicit (0+6)      &  12.3   &  15.2   &  33.8   &  35.4   &  92.4   &  22.2   &  27.0   &  34.0  \\
RC, Hybrid (3+64)       &  -5.7   &  -0.3   &   3.8   &   1.2   &  11.3   &  -0.1   &   0.6   &   1.5  \\
RC, Hybrid (3+3)        &  -4.0   &  -1.4   &   2.2   &   1.3   &  10.9   &   2.7   &   1.6   &   1.9  \\
\midrule 
CC, Explicit (3+0)      &  -11.3  &  -9.3   &  -12.0  &  -18.2  &  -8.2   &  -7.9   &  -11.1  &  -11.1 \\
CC, Implicit (0+67)     &  -22.4  &  -19.7  &  -21.5  &  -27.5  &  -25.2  &  -20.2  &  -20.9  &  -22.5 \\
CC, Implicit (0+6)      &  -20.2  &  -17.6  &  -21.5  &  -25.4  &  -23.3  &  -16.9  &  -18.0  &  -20.4 \\
CC, Hybrid (3+64)       &  -24.4  &  -20.7  &  -23.0  &  -28.4  &  -26.5  &  -20.7  &  -21.7  &  -23.6 \\
CC, Hybrid (3+3)        &  -21.7  &  -18.6  &  -22.0  &  -27.3  &  -23.8  &  -18.8  &  -19.0  &  -21.6 \\
\midrule 
CRC, Explicit (3+0)     &  -12.6  &  -11.6  &  -14.7  &  -19.2  &  -11.3  &  -10.8  &  -14.4  &  -13.5 \\
CRC, Implicit (0+67)    &  -28.5  &  -22.8  &  -24.6  &  -30.0  &  -28.6  &  -23.8  &  -25.2  &  -26.2 \\
CRC, Implicit (0+6)     &  -23.3  &  -18.4  &  -22.8  &  -26.9  &  -23.5  &  -18.2  &  -21.6  &  -22.1 \\
CRC, Hybrid (3+64)      &  -31.0  &  -25.1  &  -26.7  &  -31.8  &  -32.2  &  -25.7  &  -27.5  &  -28.6 \\
CRC, Hybrid (3+3)       &  -29.4  &  -23.3  &  -24.5  &  -28.1  &  -30.5  &  -23.5  &  -26.8  &  -26.6 \\
\midrule 
MCR, Explicit (3+0)     &  -21.9  &  -16.3  &  -18.5  &  -23.2  &  -21.0  &  -16.9  &  -20.2  &  -19.7 \\
MCR, Implicit (0+67)    &  -28.1  &  -23.3  &  -26.2  &  -30.8  &  -29.9  &  -24.5  &  -25.4  &  -26.9 \\
MCR, Implicit (0+6)     &  -24.5  &  -20.3  &  -24.0  &  -29.0  &  -24.1  &  -20.7  &  -21.2  &  -23.4 \\
MCR, Hybrid (3+64)      &  -31.3  &  -26.7  &  -28.4  &  -33.8  &  -33.4  &  -28.4  &  -28.9  &  -30.1 \\
MCR, Hybrid (3+3)       &  -29.2  &  -24.5  &  -26.0  &  -30.8  &  -32.1  &  -26.1  &  -26.4  &  -27.9 \\
\midrule
HM 16.25~\cite{hevc}    &  -32.5  &  -25.8  &  -39.4  &  -31.0  &  -45.1  &  -22.2  &  -31.9  &  -32.6 \\
\bottomrule
\end{tabular}
\vspace{-3mm}
\end{table*}

\subsection{Inter-frame Coding Frameworks}
In this work, we study four inter-frame coding frameworks: residual coding (RC), conditional coding (CC), conditional residual coding (CRC), and masked conditional residual coding (MCR). According to each framework, the inter-frame coding module with the gray background in Fig.~\ref{fig:Overview}(a) is replaced as illustrated in Fig.~\ref{fig:Overview}(b) to Fig.~\ref{fig:Overview}(e).

\textbf{Residual coding (RC):} As depicted in Fig.~\ref{fig:Overview}(b), RC encodes the residue $x_t - \ddot{x}_c$ between the coding frame $x_t$ and its pixel-domain temporal prediction $\ddot{x}_c$. Since it does not require a conditioning signal for inter-frame coding nor a mask, Condition Generator and Mask Generator in Fig.~\ref{fig:Overview}(a) are removed.

\textbf{Conditional coding (CC):} As depicted in Fig.~\ref{fig:Overview}(c), CC encodes the coding frame $x_t$ with $\dot{x}_c$ serving as the conditioning signal. Since it does not require a pixel-domain temporal prediction $\ddot{x}_c$ nor a mask, MCNet and Mask Generator in Fig.~\ref{fig:Overview}(a) are removed.

\textbf{Conditional residual coding (CRC):} As depicted in Fig.~\ref{fig:Overview}(d), CRC encodes the residue $x_t - \ddot{x}_c$ with $\dot{x}_c$ serving as the conditioning signal. Since it does not require a mask, Mask Generator in Fig.~\ref{fig:Overview}(a) is removed.

\textbf{Masked conditional residual coding (MCR):} As depicted in Fig.~\ref{fig:Overview}(a) and (e), MCR encodes the residue $x_t - m \odot \ddot{x}_c$, where $\dot{x}_c$ serves as the conditioning signal and $m$ is a pixel-wise soft mask ranging from 0 to 1. The mask $m$ is generated using the decoded flow map $\hat{f}_{t-1}$ and the pixel-domain temporal predictor $\ddot{x}_c$, following \cite{maskcrt,mmsp24,hytip}.

\subsection{Temporal Buffering Strategies}
In this work, we study three temporal buffering strategies: explicit, implicit, and hybrid buffering. The signal stored for temporal propagation in each strategy is described below:
\begin{itemize}
\item \textbf{Explicit buffering:} the decoded frame $\hat{x}_t$ as an explicit reference.
\item \textbf{Implicit buffering:} the processed latent features $\tilde{F}_t$ as an implicit reference, where $\tilde{F}_t$ is obtained by applying a $1 \times 1$ convolution to the intermediate feature $F_t$ from the decoder before reconstruction, following \cite{icme25}.
\item \textbf{Hybrid buffering:} both $\hat{x}_t$ as an explicit reference and $\tilde{F}_t$ as an implicit reference.
\end{itemize}

To analyze the impact of buffer size, we consider two settings for the implicit and hybrid variants: a large buffer with 67 channels and a small buffer with 6 channels. For explicit buffering, the buffer stores only the decoded frame, which has a fixed size of 3 channels.

\section{Experiments}
\label{sec:experiment}

\subsection{Experimental Setup}
\textbf{Training Details:}
We train our models on the Vimeo-90k dataset~\cite{vimeo}, using randomly cropped $256 \times 256$ patches. All model variants adopt the same training procedure as described in~\cite{icme25}. Following~\cite{mmsp24,icme25}, we train four models with $\lambda \in \{256, 512, 1024, 2048\}$ to target different rate-distortion trade-offs, where the hyperparameter $\lambda$ balances rate and distortion in minimizing the rate-distortion loss $\lambda D + R$. Distortion is measured as mean squared error in the RGB domain.

\textbf{Evaluation Methodologies:}
We evaluate our models on standard datasets: UVG~\cite{uvg}, HEVC Class B–E~\cite{hevcctc}, HEVC-RGB~\cite{hevcrgb}, and MCL-JCV~\cite{mcl}. Following~\cite{dcvc_dc}, YUV420 sequences are converted to RGB444 via BT.709~\cite{ffmpeg}, except HEVC-RGB. Each sequence is encoded for the first 96 frames with an intra-period of 32. Distortion and rate are measured using PSNR-RGB and bits-per-pixel (bpp), respectively. BD-rate is computed via piecewise cubic interpolation~\cite{interpolation}, where positive and negative values indicate bit rate increase and reduction, respectively. Dataset-level BD-rate is obtained by averaging per-sequence BD-rates~\cite{hevc,vvc,ecm}.

\subsection{Rate-Distortion Performance}
\label{sec:main_results}
Table~\ref{table:Main_RD_BT709} reports BD-rate results for four inter-frame coding frameworks (RC, CC, CRC, MCR) under three different buffering strategies (Explicit, Implicit, Hybrid). For implicit and hybrid buffering, we also evaluate variants with different buffer sizes. HM 16.25~\cite{hevc}, configured with the same low-delay-B setting as~\cite{hytip}, is also reported as the baseline traditional codec. The following observations are made.

\begin{figure*}[htb]
    \centering
    \includegraphics[width=1\linewidth]{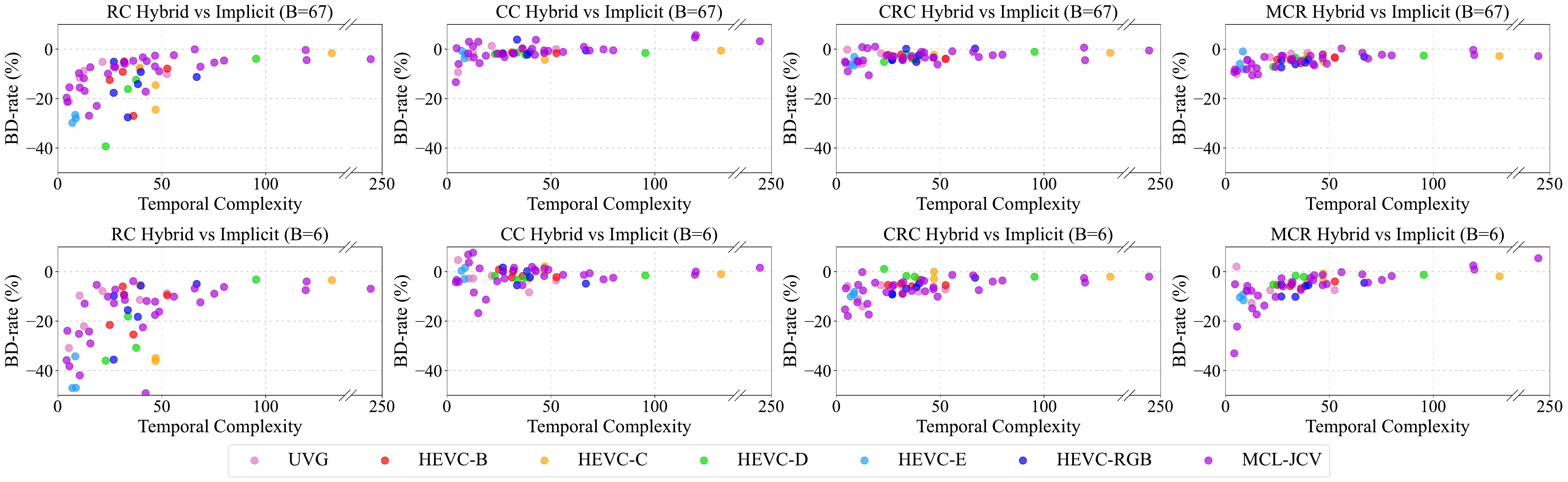}
    \vspace{-7mm}
    \caption{Per-sequence BD-rate versus temporal complexity analysis across four coding frameworks under large (B=67) and small (B=6) buffer settings. Each point represents the BD-rate of hybrid buffering using each framework's implicit buffering variant serving as the anchor. Temporal complexity is computed using the video complexity analyzer from \cite{vca}. Negative BD-rates suggest bitrate savings. The lower the BD-rate, the better the compression performance.}
    \vspace{-4mm}
    \label{fig:VCA}
\end{figure*}

(1) \textit{Effectiveness of implicit feature propagation except in residual coding}: For all frameworks except residual coding (RC), propagating intermediate features $\tilde{F}_t$ during the decoding process either solely (implicit) or together with the previously decoded frame (hybrid) outperforms using only the decoded frame (explicit). This indicates that feature propagation provides greater flexibility for temporal information propagation. RC is an exception because it encodes the residue $x_t - x_c$, so its intermediate features contain only residual information rather than rich frame content. As a result, these features provide limited contextual information for temporal prediction. This is supported by the observation that hybrid buffering, which combines explicit information, performs significantly better than implicit alone in RC. However, hybrid still performs worse than explicit in most RC cases, likely because the residual features $\tilde{F}_t$ in RC can be less effective than explicit information, $\hat{x}_{t-1}$. When the two temporal information sources are not fused well, the inclusion of $\tilde{F}_t$ may hinder performance, causing hybrid buffering to underperform compared to using explicit buffering alone. In contrast, although CRC and MCR also encode residues, their use of conditional signals enriches $\tilde{F}_t$ with contextual information, enabling hybrid and implicit buffering to surpass explicit buffering.

(2) \textit{Consistent superiority of hybrid buffering and performance hierarchy among coding frameworks}: As expected, hybrid buffering consistently outperforms implicit buffering across all coding frameworks and buffer sizes. Moreover, a clear performance hierarchy of coding frameworks is observed across buffering mechanisms, with MCR outperforming CRC, CRC outperforming CC, and CC outperforming RC. It is important to note that, as shown in Table~\ref{table:related}, prior works~\cite{crc,maskcrt,mmsp24,icme25,hytip} conduct cross-framework studies only on subsets of frameworks, and \cite{crc,maskcrt,mmsp24} limit the analysis to explicit buffering. None of these prior works offers a comprehensive comparison across both coding frameworks and buffering mechanisms as this paper. Compared to the traditional codec HM~\cite{hevc}, MCR hybrid buffering variants perform closer to HM, though a performance gap remains. This gap mainly stems from using a simple but typical learned codec as the common code base for systematic analysis. Adopting more efficient video compression architectures can further improve performance.

(3) \textit{Hybrid buffering is more robust to buffer size reduction; CC degrades the least among implicit variants}: The performance degradation when reducing buffer size is greater in implicit buffering than in hybrid buffering, with BD-rate increases of 11.7\%, 2.1\%, 4.1\%, and 3.5\% for implicit buffering, compared to 0.4\%, 2.0\%, 2.0\%, and 2.2\% for hybrid buffering in RC, CC, CRC, and MCR, respectively. This highlights the robustness of hybrid buffering under constrained memory.  Notably, conditional coding (CC) exhibits the smallest degradation (2.1\%) among all implicit buffering variants, likely because CC encodes the intra-frame $x_t$, which contains abundant content; therefore, the intermediate features $\tilde{F}_t$ during decoding naturally retain richer contextual information than those in frameworks encoding only residue ($x_t - x_c$ or $x_t - m \odot x_c$), making CC less sensitive to buffer size reduction.

(4) \textit{Increasing advantage of hybrid buffering in advanced coding frameworks}: Interestingly, as the framework transitions from CC to CRC and then MCR, the performance gap between hybrid buffering and implicit buffering becomes progressively larger. Under both large and small buffer settings, the BD-rate improvement of hybrid over implicit buffering increases from 1.1-1.2\% in CC to 2.4-4.5\% in CRC, and to 3.2-4.5\% in MCR. These results indicate that more advanced coding frameworks derive greater benefit from combining explicit and implicit temporal references, regardless of buffer size. This is because CRC (or MCR) encode the residue $x_t - x_c$ (or $x_t - m \odot x_c$), so the extracted features $\tilde{F}_t$ carry less contextual information than those derived from encoding the intra-frame $x_t$ in CC, leading to weaker performance in their implicit variants. 

\subsection{Per-Sequence Comparison}
Fig.~\ref{fig:VCA} shows how the BD-rate savings of hybrid buffering over implicit buffering across coding frameworks correlate with the temporal complexity of each test sequence. Hybrid buffering outperforms implicit buffering across most sequences and frameworks. However, some cases in CC exhibit BD-rate values greater than zero, meaning implicit buffering performs better than hybrid buffering. This may be because, as discussed in Section~\ref{sec:main_results}, the propagated feature $\tilde{F}_t$ in CC retains richer contextual information than those in RC, CRC, and MCR. When implicit and explicit information are not well fused, due to the use of a simple $1\times1$ convolution in Adapter$_P$, CC's implicit buffering alone can provide sufficient information. Nonetheless, in most CC cases, hybrid buffering still yields better results. Moreover, Table~\ref{table:Main_RD_BT709} shows that CC is not the best framework, as more advanced frameworks combined with hybrid buffering achieve superior coding performance.

In addition, hybrid buffering shows greater BD-rate savings on sequences with lower temporal complexity, likely because it uses the previously decoded frame $\hat{x}_{t-1}$, which has the highest correlation with the current coding frame $x_t$, offering a good temporal reference in lower temporal complexity scenarios.
 
\subsection{Complexity Analysis}

\begin{table}[t]
\caption{Comparison of the BD-rate and complexity in terms of the encoding/decoding MACs and model size.}
\vspace{-2mm}
\label{table:Complexity}
\centering
\begin{tabular}{l|c|ccc}
\toprule
Variants (Buffer size) & BD-rate (\%) & \begin{tabular}[c]{@{}c@{}} Enc. / Dec.\\kMACs/pixel \end{tabular} &  \begin{tabular}[c]{@{}c@{}} Model \\Size (M)  \end{tabular} \\
\midrule
RC, Explicit (3+0)     & 0       & 1029 / 661     & 7.266   \\
RC, Implicit (0+67)    & 22.3    & 1040 / 672     & 7.279   \\
RC, Hybrid (3+64)      & 1.5     & 1040 / 672     & 7.278   \\
RC, Implicit (0+6)     & 34.0    & 1029 / 661     & 7.267   \\
RC, Hybrid (3+3)       & 1.9     & 1029 / 661     & 7.267   \\
\midrule
CC, Explicit (3+0)     & -11.1   & 1162 / 768     & 7.944   \\
CC, Implicit (0+67)    & -22.5   & 1169 / 775     & 7.953   \\
CC, Hybrid (3+64)      & -23.6   & 1169 / 775     & 7.953   \\
CC, Implicit (0+6)     & -20.4   & 1162 / 768     & 7.945   \\
CC, Hybrid (3+3)       & -21.6   & 1162 / 768     & 7.945   \\
\midrule
CRC, Explicit (3+0)    & -13.5   & 1164 / 770     & 7.946   \\
CRC, Implicit (0+67)   & -26.2   & 1171 / 777     & 7.955   \\
CRC, Hybrid (3+64)     & -28.6   & 1171 / 777     & 7.955   \\
CRC, Implicit (0+6)    & -22.1   & 1164 / 770     & 7.947   \\
CRC, Hybrid (3+3)      & -26.6   & 1164 / 770     & 7.947   \\
\midrule
MCR, Explicit (3+0)    & -19.7   & 1221 / 827     & 8.169   \\
MCR, Implicit (0+67)   & -26.9   & 1228 / 834     & 8.177   \\
MCR, Hybrid (3+64)     & -30.1   & 1228 / 833     & 8.177   \\
MCR, Implicit (0+6)    & -23.4   & 1221 / 827     & 8.169   \\
MCR, Hybrid (3+3)      & -27.9   & 1221 / 827     & 8.169   \\
\bottomrule
\end{tabular}
\vspace{-4mm}
\end{table}
Table~\ref{table:Complexity} presents a detailed comparison of the BD-rate and three platform-independent complexity metrics: encoding kMACs/pixel, decoding kMACs/pixel, and model size, for different buffering strategies across four inter-frame coding frameworks. Across all coding frameworks, hybrid buffering maintains nearly the same encoding and decoding kMACs/pixel and model size to implicit buffering, with increases typically within 1\% in MACs and 0.1\% in model size compared to the explicit baseline. This minimal overhead is primarily due to the lightweight design of hybrid buffering modules, which introduces only a $1 \times 1$ convolution to adapt the buffer channel size, keeping the additional computational cost low. Notably, under small-buffer setting, hybrid buffering achieves substantial BD-rate improvements of 10.5\%/1.2\% in CC, 13.1\%/4.5\% in CRC, and 8.2\%/4.5\% in MCR over their respective explicit/implicit variants, while maintaining nearly the same MACs and model size. This favorable performance–complexity trade-off highlights the practicality of hybrid buffering, especially in resource-constrained scenarios.

\section{Conclusion}
\label{sec:conclusion}
This paper systematically evaluates explicit, implicit, and hybrid temporal buffering strategies across four inter-frame coding frameworks under unified settings. Results show hybrid buffering consistently outperforms explicit buffering with minimal added complexity. Advanced frameworks like conditional residual coding and masked conditional residual coding benefit most from hybrid buffering through effective use of explicit and implicit information. These findings offer clear insights into how buffering strategies affect coding performance and complexity under unified settings, supporting informed design choices for future learned video codecs, especially in memory-constrained scenarios. Expanding the analysis by adopting more advanced compression architectures in the unified codec to examine different coding frameworks and buffering strategies under similar model complexity is among our future work.

\bibliography{references}
\bibliographystyle{IEEEtran}


\end{document}